\documentclass[a4paper,12pt]{article}
\usepackage[utf8]{inputenc}

\usepackage[left=2.5cm,right=2.5cm,bottom=4cm]{geometry}

\usepackage{mathtools,amsmath,amssymb}
\usepackage{graphicx}

\numberwithin{equation}{section} 

\usepackage{lmodern}
\usepackage[T1]{fontenc}
\usepackage{microtype}
\usepackage{cite}

\usepackage{hyperref}

\newcommand{\half}{{\frac{1}{2}}}

\newcommand{\A}{{\rm A}}
\newcommand{\B}{{\rm B}}
\newcommand{\C}{{\rm C}}
\newcommand{\D}{{\rm D}}
\newcommand{\X}{{\rm X}}

\newcommand{\osc}[1]{\boldsymbol{\mathrm{#1}}}

\newcommand{\oa}{\osc{a}}

\newcommand{\oad}{\osc{\bar{a}}}

\newcommand{\spec}{y}

\newcommand{\ro}{\mathcal{R}}
\newcommand{\qop}{\mathbf{Q}}

\newcommand{\mqm}{\mathbf{M}_{-}}
\newcommand{\mq}{\mathbf{M}}

\newcommand{\m}{m}

\DeclareMathOperator{\tr}{tr}

\author{Rouven Frassek}
\title{Algebraic Bethe ansatz for Q-operators}

\begin{document}

\begin{titlepage}\strut\hfill DCPT-15/17

\vspace{.5in}

\begin{center}
 \textbf{\LARGE Algebraic Bethe ansatz for Q-operators:\\[5mm]
The Heisenberg spin chain}
\\\vspace{1in}
\large Rouven Frassek 
\\[0.2in]
\texttt{
      \href{mailto:rouven.frassek@durham.ac.uk}{rouven.frassek@durham.ac.uk}
    }
\\[0.6in]
\large Department of Mathematical Sciences, Durham University,\\
     South Road, Durham DH1 3LE, United Kingdom
 \end{center}
 \vspace{.8in}
\begin{center}
\textbf{\large Abstract}
\end{center}
\begin{center}
\begin{minipage}{400pt}
\noindent  We diagonalize Q-operators for rational homogeneous $\mathfrak{sl}(2)$-invariant Heisenberg spin chains using the algebraic Bethe ansatz. After deriving the fundamental commutation relations relevant for this case from the Yang-Baxter equation we demonstrate that the Q-operators act diagonally on the Bethe vectors if the Bethe equations are satisfied. In this way we provide a direct proof that the eigenvalues of the Q-operators studied here are given by Baxter's Q-functions. 
\end{minipage}
\end{center}
\end{titlepage}

\vfill
\newpage

\section{Introduction}
Integrable spin chains are prominent examples of integrable models. The solution of the Heisenberg $\text{XXX}_\frac{1}{2}$ spin chain goes back to Hans Bethe~\cite{Bethe1931}. His ansatz of factorized plane wave scattering is today known as the coordinate Bethe ansatz and is probably the most intuitive way to solve the eigenvalue problem. However, the origin of the integrable structure remains unclear. 

The situation is somehow different when studying integrable models using the quantum inverse scattering method which, in particular, provides a prescription of constructing commuting families of operators. Combined with the algebraic Bethe ansatz, that allows to determine the wave function for models with a suitable reference state, it yields a powerful tool  to diagonalize integrable Hamiltonians along with the commuting family of operators, see e.g.~\cite{Faddeev2007}. 

Q-operators are distinguished members of the family of commuting operators. Their eigenvalues,  so-called Baxter's Q-functions, are in general given by polynomials with zeros at the Bethe roots~\cite{Baxter2007}. They are of fundamental importance for the so-called fermionic basis, see e.g.~\cite{Boos2008}, as well as in the {\small ODE/IM} correspondence~\cite{Dorey:2000ma}. Furthermore, the spectral problem of planar $\mathcal{N}=4$ super Yang-Mills theory in four dimensions, which is believed to be exactly solvable in the planar limit~\cite{Beisert2010a}, resulted in a set of relations among Q-functions, see~\cite{Gromov2014a} and references therein. These all-loop equations are the same as for the spin chain, a close cousin of the Heisenberg chain studied here, that appears in the spectral problem at one-loop level.
Initially Q-operators were introduced by Baxter to solve the eight-vertex model without specifying a reference state~\cite{Baxter2007}. However, the limit to the rational case is singular. 

Inspired by~\cite{Bazhanov1999}, Q-operators were built in the framework of the quantum inverse scattering method for finite-dimensional rational spin chains in~\cite{Bazhanov2010a,Bazhanov2010,Frassek2011}. The construction naturally incorporates a magnetic field $\phi$ that manifests itself as a diagonal twist in the transfer matrices, see also~\cite{Korff2006}. The twist field lifts the usual degeneracy in the spectrum and allows in the case of $\mathfrak{sl}(2)$  for the definition of two Q-operators $\qop_\pm$ which belong to the commuting family of operators. A main ingredient in the construction~\cite{Bazhanov2010a,Bazhanov2010,Frassek2011} are certain factorization formulas that relate the Q-operators to transfer matrices and yield the quantum Wronskian on an operatorial level. Having derived the quantum Wronskian and in particular the QQ-relations on the operatorial level it was argued that the eigenvalues of the constructed Q-operators are indeed the Q-functions, see Section~3.5 in~\cite{Bazhanov2010a}. Furthermore, it was shown in~\cite{Frassek2013} how local charges can be obtained from the Q-operators. For a discussion of Q-operators for supersymmetric spin chains relevant in the spectral problem, see~\cite{Frassek2010,carlophd,Frassek2014}. 

In the following, we diagonalize the Q-operators of the length $N$ Heisenberg $\text{XXX}_s$ spin chain introduced in~\cite{Bazhanov2010a,Frassek2011} directly using the algebraic Bethe ansatz. In analogy to the algebraic Bethe ansatz for the corresponding transfer matrix we define the off-shell Bethe vectors as the $m$-fold excitations on the tensor product of highest or lowest weight states. In order to evaluate the action of the Q-operators on the off-shell Bethe vectors we derive certain fundamental commutation relations from the Yang-Baxter equation. We show that the Q-operators act diagonally on the Bethe vectors
\begin{equation}\label{qpsi}
 \qop_\pm(y)\vert \psi^\pm_\m \rangle =Q_\pm(y)\vert \psi^\pm_\m \rangle\,,
\end{equation} 
with the spectral parameter $y$ and Baxter's Q-functions
\begin{equation}\label{qfunk}
 Q_\pm(y)=e^{\pm i y\phi}\prod_{j=1}^{\m}(y-z_j^\pm)\,,
\end{equation} 
if the Bethe roots $z_k^\pm$ satisfy the Bethe equations
\begin{equation}\label{baein2}
\left(\frac{z_k^\pm+s}{z_k^\pm-s}\right)^N=e^{\pm 2i\phi}\prod_{\substack{j=1\\j\neq k}}^m\frac{z_k^\pm-z_j^\pm+1}{z_k^\pm-z_j^\pm-1}\,.
\end{equation}  
This pragmatic approach motivates the choice of Lax operators used to construct the Q-operators, avoids the detour of deriving the functional relations from the factorization formulas and provides a transparent derivation of the eigenvalues of the Q-operators along with the corresponding eigenvectors. 

This article is organized as follows: 
In Section~\ref{qops} we provide a short review of the Q-operator construction as presented in~\cite{Bazhanov2010a,Frassek2011}. The elementary solutions used to construct the Q-operators satisfy certain Yang-Baxter relations. This is discussed in Section~\ref{ybeee}. In particular, the Yang-Baxter equation relevant to derive the fundamental commutation relations in Section~\ref{sec:fcr} is derived. Section~\ref{fq2q} contains the algebraic Bethe ansatz for one of the Q-operators of the  Heisenberg $\text{XXX}_s$ spin chain. We show that the Q-operator $\qop_+$ acts diagonally on the Bethe states if the Bethe equations are satisfied. Finally, we use a relation between the Q-operators to obtain the eigenvalues and eigenfunctions of  $\qop_-$. Further material is collected in the appendices. 

\section{Q-operators for the Heisenberg spin chain}\label{qops}
In this section we construct the Q-operators for the Heisenberg $\text{XXX}_s$ spin chain following~\cite{Bazhanov2010a,Frassek2011}. The elementary Lax operators used to construct Q-operators of the twisted Heisenberg spin chain can conveniently be written as
\begin{equation}\label{rop}
 \mathcal{R}_{s,\pm}(\spec)=e^{\oad S_\mp}\,\ro_{\pm}^0(\spec)\,e^{-\oa S_\pm}\,,
\end{equation} 
with the oscillator algebra and the $\mathfrak{sl}(2)$ generators obeying the commutation relations
\begin{equation}\label{gens}
 [\oa,\oad]=1\,,\quad \quad  [S_+,S_-]=2S_3\,,\quad\quad [S_3,S_\pm]=\pm S_\pm\,.
\end{equation} 
The diagonal part $\ro_{\pm}^0(\spec)$ depends on the spectral parameter $\spec$ and  reads
\begin{equation}\label{normr}
\ro_{\pm}^0(\spec)=\frac{\Gamma(\spec\mp S_3)}{\Gamma\left(\spec-s\right)}\,.
\end{equation} 
Here we have chosen a suitable normalization for highest/lowest weight representations, i.e. for
\begin{equation}\label{hwss}
 S_\pm\vert \omega^\pm\rangle=0\,,\quad\quad  S_3\vert \omega^\pm\rangle=\pm s\vert \omega^\pm\rangle\,,
\end{equation} 
we have $\ro_{\pm}^0(\spec)\vert \omega^\pm\rangle=\vert\omega^\pm\rangle$.
This normalization ensures that all entries of the Q-operators will be polynomials in the spectral parameter when considering finite-dimensional representations $s=\frac{1}{2},1,\frac{3}{2},\ldots$ up to an overall factor, see below. The Lax operators \eqref{rop} are also employed in the so-called {\small DST}-chain, see e.g.~\cite{kovalsky}. However, here the role of the auxiliary and quantum space is interchanged. The auxiliary space of the monodromy of the Q-operators for rational spin chains is built from the product in the oscillator space and from the $N$-fold tensor product in the $\mathfrak{sl}(2)$ space
\begin{equation}\label{qmon}
 \mq_\pm(\spec)= \underbrace{\mathcal{R}_{s,\pm}(\spec)\otimes \mathcal{R}_{s,\pm}(\spec)\otimes\ldots\otimes  \mathcal{R}_{s,\pm}(\spec)}_N\,.
\end{equation} 
Here we used $\otimes$ to denote the tensor product in the first space and multiplication in the second space of $\mathcal{R}_{s,\pm}$.
Following~\cite{Bazhanov2010a,Frassek2011}, we define the Q-operators as the regulated trace of the monodromy \eqref{qmon}
\begin{equation}\label{qop}
 \qop_\pm(\spec)=e^{\pm i\spec\phi}Z_\pm^{-1} \tr\left[ e^{\mp 2i\phi \oad\oa}\,\mq_\pm(\spec)\right]\,,
\end{equation} 
where the trace sums over all states in the Hilbert space and we introduced the normalization
\begin{equation}
 Z_\pm=\tr\left[ e^{\mp 2i\phi \oad\oa}\right]=\frac{1}{1-e^{\mp 2i\phi}}\,.
\end{equation} 
The parameter $\phi$ is the twist angle and manifests itself as a diagonal twist in the transfer matrices. It is known that such diagonal twisted boundary conditions break the $\mathfrak{sl}(2)$ symmetry but the quantum number $m$ is still well defined. As a consequence, not only highest weight states but also all decendents can be obtained from the Bethe ansatz. 

\section{Yang-Baxter relations}\label{ybeee}
The fundamental relation underlying the quantum inverse scattering method and the algebraic Bethe ansatz for rational spin chains is the Yang-Baxter equation. It reads
\begin{equation}\label{rttttt}
 R_{1,2}(x-y)R_{1,3}(x)R_{2,3}(y)=R_{2,3}(y)R_{1,3}(x) R_{1,2}(x-y)\,,
\end{equation} 
and is defined on the tensor product of three vector spaces $V_1\otimes V_2\otimes V_3$.
Here the R-matrix $R_{i,j}$ acts trivially as the identity on $V_k$ with $k\neq i,j$ and non-trivially on $V_i\otimes V_j$. For further references we refer the reader to the excellent introduction~\cite{Faddeev2007}.

The Lax operators $\mathcal{L}$ relevant to construct the transfer matrix of the Heisenberg chain with the fundamental representation in the auxiliary space satisfy the {\small RTT}-relation which serves as a defining relation of the Yangian, see e.g.~\cite{molevbook}. Namely, it holds that
\begin{equation}\label{rttt}
 \mathbf{R}_{\square,\square}(x-y)\mathcal{L}_{\square,s}(x)\mathcal{L}_{\square,s}(y)=\mathcal{L}_{\square,s}(y)\mathcal{L}_{\square,s}(x) \mathbf{R}_{\square,\square}(x-y)\,.
\end{equation} 
Here the fundamental R-matrix reads
\begin{equation}\label{rmatrix}
 \mathbf{R}_{\square,\square}(x)=\left(\begin{array}{cccc}
              x+1&0&0&0\\
              0&x&1&0\\
              0&1&x&0\\
              0&0&0&x+1
             \end{array}\right)\,,
\end{equation} 
and acts trivially on the third space denoted by $s$ in \eqref{rttt}. The $\mathfrak{sl}(2)$-invariant Lax operator can be written as
\begin{equation}\label{laxi}
 \mathcal{L}_{\square,s}(y)=\left(\begin{array}{cc}
              y+1+S_3&S_-\\
	      S_+&y+1-S_3
             \end{array}\right)\,,
\end{equation} 
with the generators of $\mathfrak{sl}(2)$ defined in \eqref{gens}.
Here, the shift in the spectral parameter has been introduced for convenience. Its role will become clear later on. Before presenting the Yang-Baxter equation that is relevant to derive the fundamental commutation relations necessary to diagonalize the Q-operators in \eqref{qop} we recall the derivation of the Lax operators $\mathcal{R}_{s,\pm}$ \eqref{rop} employed in the construction~\cite{Bazhanov2010a,Frassek2011}. Apart from the solutions that are classified by the Yangian, the {\small RTT}-relation allows for solutions where the dependence on the spectral parameter is not  proportional to the diagonal, cf.~\cite{Bazhanov2010} as well as~\cite{Rolph2014} where these solutions were studied in the context of Drinfeld's second realization. In particular, it holds that
\begin{equation}
 \mathbf{R}_{\square,\square}(x-y)L_{\square,\pm}(x)L_{\square,\pm}(y)=L_{\square,\pm}(y)L_{\square,\pm}(x) \mathbf{R}_{\square,\square}(x-y)\,,
\end{equation} 
with
\begin{equation}\label{plax}
L_{\,\square,+}(z)=\left(\begin{array}{cc}
              1&-\oa\\
	      \oad&z-\oad\oa
             \end{array}\right)
\quad  \text{and} \quad 
L_{\,\square,-}(z)=\left(\begin{array}{cc}
              z-\oad\oa&\oad\\
	      -\oa&1
             \end{array}\right) \,.
\end{equation}
These solutions where used for the Q-operator construction in~\cite{Bazhanov2010a}. Furthermore, they are relevant to obtain the Lax operators for Q-operators for arbitrary representations of $\mathfrak{sl}(2)$ in the quantum space in~\cite{Frassek2011}, see also~\cite{carlophd}. Another ingredient is the $\mathfrak{sl}(2)$-invariant Lax operator for the transfer matrix of rational spin chains \eqref{laxi}.
The relevant Yang-Baxter relation that defines the Lax operators $\mathcal{R}_{s,\pm}$ in \eqref{rop} reads
\begin{equation}\label{ybealt}
 \mathcal{L}_{\square,s}(x-y)L_{\,\square,\pm}(x)\mathcal{R}_{s,\pm}(y)=\mathcal{R}_{s,\pm}(y)L_{\,\square,\pm}(x) \mathcal{L}_{\square,s}(x-y)\,.
\end{equation} 
We note that for spin $\half$ we identify $\mathcal{R}_{\half,\pm}(z+\frac{1}{2})=L_{\square,\pm}(z)$. However, this relation is not suitable for our purposes. 

To obtain the fundamental commutation relations that are needed to diagonalize Q-operators we define the Lax operator 
\begin{equation}\label{laxbar}
 \mathcal{L}_{s,\square}(z)=\left(\begin{array}{cc}
              z+S_3&S_-\\
	      S_+&z-S_3
             \end{array}\right)\,.
\end{equation} 
It satisfies the unitarity relation
\begin{equation}\label{unitlax}
 \mathcal{L}_{s,\square}(x-y) \mathcal{L}_{\square,s}(y-x)=(y-x)(x-y-1)+\frac{c}{2}\,,
\end{equation} 
where $c$ is the Casimir element of $\mathfrak{sl}(2)$ given by  $c=S_+\,S_-+S_-\,S_++2S_3\,S_3$. We note that the Lax operator \eqref{laxbar} satisfying \eqref{unitlax} only differs by a shift in the spectral parameter from the Lax operator defined in \eqref{laxi}. This is not always the case and for convenience we distinguish between the two solutions \eqref{laxi} and \eqref{laxbar}. In addition to the Lax operators $L_{\square,\pm}$ in \eqref{plax} we define
\begin{equation}\label{plaxbar}
L_{+,\square}(z)=\left(\begin{array}{cc}
              z+1+\oad\oa&-\oa\\
	      \oad&-1
             \end{array}\right)\quad  \text{and} \quad L_{-,\square}(z)=\left(\begin{array}{cc}
              -1&\oad\\
	      -\oa&z+1+\oad\oa
             \end{array}\right) \,,
\end{equation}
that obey the unitarity conditions
\begin{equation}\label{unitplax}
L_{+,\square}(x-y)L_{\square,+}(y-x)=x-y\,,\quad\quad  L_{-,\square}(x-y)L_{\square,-}(y-x)=x-y \,.
\end{equation} 
In contrast to the Lax operators $\mathcal{L}$ these solutions significantly differ from \eqref{plax}. The Yang-Baxter equation relevant to derive the sought-after fundamental commutation relations can then be obtained from \eqref{ybealt} after multiplying \eqref{laxbar} and \eqref{plaxbar}. We find
\begin{equation}\label{fcrybe}
  \mathcal{R}_{s,\pm}(y)\mathcal{L}_{s,\square}(x)L_{\pm,\square}(x-y)=L_{\pm,\square}(x-y) \mathcal{L}_{s,\square}(x)\mathcal{R}_{s,\pm}(y)\,,
\end{equation} 
where we used the unitarity relations \eqref{unitlax} and \eqref{unitplax} and for convenience  renamed the spectral parameters.

\section{Fundamental commutation relations}\label{sec:fcr}
 Following the algebraic Bethe ansatz we define the monodromy of the \textit{fundamental transfer matrix}\,\footnote{In contrast to the convention often found in the literature where the fundamental transfer matrix refers to the transfer matrix of equal representation in quantum and auxiliary space.} as the matrix product of the Lax operators introduced in  \eqref{laxbar} in the two dimensional auxiliary space and the $N$-fold tensor product in the quantum space\,\footnote{Note that the monodromy in \eqref{transmon} is related to one built from the Lax operator in \eqref{laxi} by a shift in the spectral parameter, cf. \eqref{laxbar}.} labelled by the representation label $s$. 
 Using the same conventions as in the definition of the monodromy of the Q-operators \eqref{qmon} it can be written as
 \begin{equation}\label{transmon}
  \mathcal{M}(y)=\mathcal{L}_{s,\square}(y)\otimes\mathcal{L}_{s,\square}(y)\otimes\ldots\otimes\mathcal{L}_{s,\square}(y)\,.
 \end{equation} 
 By construction $\mathcal{M}$ is a $2\times 2$ matrix in the auxiliary space with operatorial entries built from the $\mathfrak{sl}(2)$ generators at the different sites
 \begin{equation}\label{monab}
   \mathcal{M}(y) =\left(\begin{array}{cc}
              \A(y)&\B(y)\\
	      \C(y)&\D(y)
             \end{array}\right)\,.
 \end{equation}
  The corresponding transfer matrix is constructed as the trace over the two-dimensional auxiliary space containing a suitable boundary twist that acts trivially in the quantum space
 \begin{equation}\label{transi}
  \mathbf{T}_\square(y)=\tr\,e^{-2 i\phi S_3}  \mathcal{M}(y)=e^{-i\phi}\A(y)+e^{+i\phi}\D(y)\,.
 \end{equation} 
 As a consequence of the Yang-Baxter relation \eqref{rttt} and the unitarity relation \eqref{unitlax} the monodromy $\mathcal{M}$ in \eqref{transmon} satisfies the {\small RTT}-relation
 \begin{equation}\label{RRRRETTT}
 \mathbf{R}_{\square,\square}(x-y)(\mathbb{I}\otimes\mathcal{M}(x))(\mathcal{M}(y)\otimes\mathbb{I})=(\mathcal{M}(y)\otimes\mathbb{I})(\mathbb{I}\otimes\mathcal{M}(x)) \mathbf{R}_{\square,\square}(x-y)\,,
 \end{equation} 
where $\mathbb{I}$ denotes the $2\times 2$ identity matrix. To obtain the  fundamental commutation relations we insert the explicit form of the R-matrix $\mathbf{R}_{\square,\square}$  as given in \eqref{rmatrix} into \eqref{RRRRETTT}, see e.g.~\cite{Faddeev2007,Slavnov2007}. In particular, we find
\begin{gather}
 \B(x)\B(y)=\B(y)\B(x)\,,\label{Bcom}\\
 \nonumber\\
 \A(x)\B(y)=f(y,x)\,\B(y)\A(x)-g(y,x)\,\B(x)\A(y)\,,\label{abcom}\\
 \nonumber\\
 \D(x)\B(y)=f(x,y)\,\B(y)\D(x)-g(x,y)\,\B(x)\D(y)\,,\label{dbcom}
 \\
 \nonumber\\
 \C(x)\B(y)=\B(y)\C(x)+g(x,y)\left[\,\A(y)\D(x)-\,\A(x)\D(y)\right]\,.\label{cbcom}
\end{gather}
Here we used the notation
\begin{equation}\label{fg}
 f(x,y)=\frac{1+x-y}{x-y}\quad \text{and}\quad  g(x,y)=\frac{1}{x-y}\,,
\end{equation} 
with $ f(x,y)=g(x,y)+1$.

To diagonalize the Q-operators in \eqref{qop} we derive the commutation relations of the Q-operator's monodromy $\mq_\pm$ with the operators $\A$, $\B$, $\C$ and $\D$ in \eqref{monab}. As a consequence of the Yang-Baxter relation \eqref{fcrybe}, we find that the monodromy of the Q-operators \eqref{qmon} and the monodromy of the fundamental transfer matrix \eqref{transmon} satisfy
\begin{equation}\label{fcrrtt}
  \mq_\pm(y)\,\mathcal{M}(x)\,L_{\pm,\square}(x-y)=L_{\pm,\square}(x-y)\, \mathcal{M}(x)\,\mq_\pm(y)\,.
\end{equation} 
In the following we focus on the Q-operator $\mathbf{Q}_+$ built from the monodromy $\mq_+$. In this case the additional fundamental commutation relations necessary to diagonalize the Q-operator $\mathbf{Q}_+$ arise from \eqref{fcrrtt} after inserting the explicit form of $L_{+,\square}$ as given in \eqref{plaxbar}. We obtain
\begin{gather}
 \mq_+(y)[(x-y+1+\oad\oa)\A(x)+\oad\,\B(x)]=[\A(x)(x-y+1+\oad\oa)-\C(x)\,\oa ]\,\mq_+(y)\,,\label{amm}\\
 \nonumber\\
  \mq_+(y)\B(x)=\B(x)(y-x-1-\oad\oa)\,\mq_+(y)+\D(x)\,\oa\, \mq_+(y)-\mq_+(y)\,\oa\,\A(x)\,,\label{mb}\\
  \nonumber\\
   \C(x)\,\mq_+(y)=\mq_+(y)(y-x-1-\oad\oa)\C(x)+\A(x)\,\oad\, \mq_+(y)-\mq_+(y)\,\oad\,\D(x)\,,\label{cmm}\\
   \nonumber\\
    [\D(x),\mq_+(y)]=\B(x)\,\oad\, \mq_+(y)+\mq_+(y)\,\oa\,\C(x)\,.\label{bd}
\end{gather} 
For our purposes it is enough to consider \eqref{mb} and \eqref{bd} and combine them into the relation
\begin{equation}\label{combined}
 \mq_+(y)\B(x)=(y-x)\B(x)\,\mq_+(y)+\X(y,x)\,,
\end{equation} 
with 
\begin{equation}\label{xk}
 \X(y,x) =\oa\, \mq_+(y) \D(x)+\oa\, \mq_+(y)\,\oa\, \C(x)-\mq_+(y)\,\oa\,\A(x)\,.
\end{equation} 
The relations relevant to diagonalize $\mathbf{Q}_-$ involving $\mq_-$ and $L_{-,\square}$ can be found in Appendix~\ref{fcrsq2}. Similar relations appeared previously for the product of two Q-operators in~\cite{Korff2006}. 
\section{From Q-operators to Q-functions}\label{fq2q}
In this section we diagonalize the Q-operator $\qop_+$ following the algebraic Bethe ansatz. In addition to the fundamental commutation relations \eqref{Bcom}--\eqref{cbcom} we will repeatedly use the commutation relation among the monodromy $\mq_+$ and the B-operators \eqref{combined} that we obtained from the Yang-Baxter equation \eqref{fcrrtt}.

Following~\cite{Faddeev2007}, we introduce the reference state
\begin{equation}\label{vac}
 \vert\Omega^+\rangle=\underbrace{\vert\omega^+\rangle\otimes\ldots\otimes\vert\omega^+\rangle}_N\,,
\end{equation} 
with the highest weight state at each site as defined in \eqref{hwss}.
The action of the monodromy \eqref{transmon} on the reference state $\vert\Omega^+\rangle$ yields in particular
\begin{equation}\label{ADC}
 \A(y)\vert\Omega^+\rangle=\alpha(y)\vert\Omega^+\rangle\,,\quad  \D(y)\vert\Omega^+\rangle=\delta(y)\vert\Omega^+\rangle\,,\quad \C(y)\vert\Omega^+\rangle=0\,,
\end{equation} 
with
\begin{equation}\label{ad}
 \alpha(y)=\left(y+s\right)^N\,, \quad  \delta(y)=\left(y-s\right)^N\,.
\end{equation} 
Due to our choice of normalization of the operators $\mathcal{R}_\pm$ in \eqref{normr} the action of the monodromy \eqref{qmon} of the Q-operator $\qop_+$  on the reference state $\vert\Omega^+\rangle$ is independent of the spectral parameter $y$ and reads 
\begin{equation}\label{qmonvac}
 \mq_+(y)\vert \Omega^+\rangle=e^{\oad S_-^\text{tot}}\vert \Omega^+\rangle\,,
\end{equation} 
where $S_-^\text{tot}=\sum_{i=1}^N S_-^{(i)}$ with $S_-^{(i)}$ denoting the generator $S_-$ acting on site $i$. The off-shell Bethe sates in the context of the algebraic Bethe ansatz are then defined as
\begin{equation}\label{offshell}
\vert \psi_\m^+\rangle=\B(z_1^+)\B(z_2^+)\cdots \B(z_\m^+)\vert\Omega^+\rangle\,.
\end{equation} 
The states $\vert\psi_\m^+\rangle$ are eigenstates of the corresponding transfer matrix if the parameters $z_i$ satisfy the Bethe equations~\cite{Faddeev2007}. Thus we expect that the Q-operators  $\qop_\pm$ also act diagonally as they belong to the family of commuting operators and obey the Baxter equation, see~\cite{Bazhanov2010a}.

Instead of studying the action of the Q-operator on the off-shell Bethe states \eqref{offshell} it is convenient to act with the monodromy and take the trace afterwards. Using the fundamental commutation relations in \eqref{combined} we obtain
 \begin{equation}\label{actqmon}
\mq_+(y)\vert \psi^+_\m\rangle=\prod_{i=1}^\m (y-z_i^+)\B(z_i^+)\mq_+(y)\vert\Omega^+\rangle+\vert y,\{z^+\} \rangle\,,
\end{equation} 
where
 \begin{equation}\label{unwanmon}
 \begin{split}
\vert y,\{z^+\} \rangle=\sum_{k=1}^\m \prod_{i=1}^{k-1}(y-z_i^+)\B(z_1^+)\cdots\B(z_{k-1}^+)\,\X(y,z_k^+) \B(z_{k+1}^+)\cdots\B(z_{\m}^+)\vert\Omega^+\rangle\,.
 \end{split}
\end{equation} 
Here, the operator $\X$ is defined in \eqref{xk} and $\{z^+\}$ denotes the set of parameters $z_1^+,\ldots,z_\m^+$.
The role of the terms on the right-hand-side of \eqref{actqmon} becomes clear after we have  taken the trace over the auxiliary space and using the action of the monodromy $\mq_+$ on the reference state \eqref{qmonvac}. From the definition of the Q-operators \eqref{qop} we find that
\begin{equation}\label{qactt}
 \qop_+(y)\vert \psi^+_\m\rangle=e^{+iy\phi}\prod_{i=1}^\m (y-z_i^+)\vert \psi^+_\m\rangle+\vert y,\{z^+\} \rangle_\phi\,,
\end{equation} 
with 
\begin{equation}\label{unwanq}
 \vert y,\{z^+\} \rangle_\phi= e^{+ iy\phi}Z_+^{-1} \tr \left[e^{-2i\phi \oad\oa}\,\vert y,\{z^+\} \rangle\right]\,.
\end{equation} 
The term proportional to $\vert\psi^+_\m\rangle$ yields the Q-function $Q_+$ if the parameters $z_i^+$ are Bethe roots, see~\eqref{qfunk}. Thus, we expect that the \textit{unwanted terms} $ \vert y,\{z^+\} \rangle_\phi$ vanish if the parameters $z_i^+$ satisfy the Bethe equations \eqref{baein2}.

 From the commutation relations among the B-operators \eqref{Bcom} it is obvious that the left-hand-side of \eqref{actqmon} is symmetric in the parameters $z_i$ as well as the first term on the right-hand-side. As a consequence also $\vert y,\{z^+\}\rangle$ and $\vert y,\{z^+\}\rangle_\phi$ given in \eqref{unwanmon} and \eqref{unwanq} must be invariant under permutations of $z_i^+$. This property will be important later.

The evaluation of the unwanted terms \eqref{unwanq} is more involved than in the algebraic Bethe ansatz for the corresponding transfer matrix.
The reason is that while the operators $\A$ and $\D$ conserve the number of $\B$-operators, the operators $\mq_+$ and $\C$ can annihilate one B-operator, see \eqref{abcom}--\eqref{cbcom} and \eqref{combined}. After we have commuted all operators $\A$, $\D$, $\C$ and $\mq_+$ through the B-operators and evaluating them on the reference state as discussed \eqref{ADC} and \eqref{qmonvac} the unwanted terms in \eqref{unwanq} can be written as
\begin{equation}\label{coeffs}
\begin{split}
 \vert y,\{z^+ \} \rangle_\phi= e^{+iy\phi}\sum_{\sigma\in\mathcal{S}_\m }\sum_{k=1}^m\frac{\mathcal{F}_k^{\,\phi}(y,z_{\sigma(1)}^+,\ldots,z_{\sigma(m)}^+)}{(e^{2i\phi}-1)^k} \B(z_{\sigma(k+1)}^+)\cdots\B(z_{\sigma(m)}^+)\, S^k_{\text{tot}}\, \vert \Omega^+ \rangle\,.
  \end{split}
\end{equation} 
Here we moved the oscillators multiplied to the left of the monodromy $\mq_+$ to the right by commuting them through the regulator $e^{-2i\phi\oad\oa}$ and subsequently using the cyclicity of the trace. Finally, we can take the trace of $\mq_+$ on the reference state
\begin{equation}\label{traceev}
\tr e^{-2i\phi\oad\oa}\,\mq_+(y)\oa^k\vert\Omega^+\rangle
=\frac{Z_+}{(e^{2i\phi}-1)^{k}}S^k_{\text{tot}}\vert\Omega^+\rangle\,.
\end{equation} 

Assuming that the states corresponding to different families $\{z^+\}$ are linearly independent, see also the discussion in~\cite{Slavnov2007},  we demand that all coefficients in \eqref{coeffs} vanish
\begin{equation}\label{cond1}
 \mathcal{F}_k^{\,\phi}(y,z_{\sigma(1)}^+,\ldots,z_{\sigma(m)}^+)=0\,,
\end{equation} 
for all permutations $\sigma\in\mathcal{S}_\m$. In Section~\ref{sec:first} we evaluate $\mathcal{F}_1^{\,\phi}$ for which the condition \eqref{cond1} yields the Bethe equations \eqref{baein2} and in Section~\ref{sec:second} we show that all other $\mathcal{F}_k^{\,\phi}$ with $k>1$ vanish if the Bethe equations are satisfied. 

There exist exceptional solutions to the Bethe equations which in particular include coinciding Bethe roots or so called exact string solutions which may yield null vectors or singular solutions, see e.g.~\cite{Hao2013,Vieira2015} and references therein. To some extend the situation is more feasable in the case of twisted boundary conditions. However, in the following we focus on  the generic case where $z_i^+\neq z_j^+$ and $z_i^+-z_j^+\neq\pm 1$.

\subsection{Bethe equations: The first term}\label{sec:first}
Having in mind that the unwanted terms in \eqref{unwanmon} are symmetric in the parameters $z_i^+$ we calculate the first term $\mathcal{F}_1^{\,\phi}$ employing an argument similar to the one used to obtain the unwanted terms for the corresponding spin chain transfer matrix, see~\cite{Faddeev2007,Slavnov2007}. Let us consider an ordering of the parameters $z_i^+$ as in \eqref{unwanmon} and focus on the first term in the sum where $k=1$.
In particular, all unwanted terms \eqref{coeffs} that are independent of $\B(z_1^+)$ must arise from it after taking the trace! We denote the equality of the part independent  of $\B(z_1^+)$ as $\simeq$ such that 
\begin{equation}\label{eqrel}
 \vert y,\{z^+\}\rangle\simeq  \X(y,z_1^+)\vert I\rangle\quad\quad\text{with}\quad\quad \vert {I}\rangle=\prod_{k\in I}B(z_{k}^+)\vert \Omega \rangle\,,
\end{equation} 
as discussed above. Here the set $I$ is given by $I=\{2,3,\ldots,\m\}$. 

Let us evaluate the action of the operators $\A$, $\D$ and $\C$ on $\vert I\rangle$ in $\X(y,z_1^+)\vert I\rangle$, cf.~\eqref{xk}. This can be done by repeatedly applying the fundamental commutation relations \eqref{abcom}, \eqref{dbcom} and \eqref{combined} and evaluate the relevant operators on the reference state \eqref{ADC}. The action of the operators $\A$, $\D$ and $\C$ on $\vert I\rangle$ is well-known, see e.g.~\cite{Slavnov2007}, and can be found in \eqref{actaa}--\eqref{actcc}. We obtain
\begin{equation}\label{actx}
\begin{split}
  \X(y,z_1^+)\vert I\rangle&=\big(\mathcal{D}^I(z_1^+)\,\oa\, \mq_+(y)-\mathcal{A}^I(z_1^+)\, \mq_+(y)\, \oa\big)\,\vert {I}\rangle\\
  &\quad+\sum_{j\in {I}}\mathcal{C}^{I}_j(z_1^+)\,\oa\,\mq_+(y)\, \oa\,\vert {{I}\setminus \{j\}}\rangle\\
  &\quad+\sum_{j\in I}\big(\mathcal{D}^I_j(z_1^+)\,\oa\, \mq_+-\mathcal{A}^I_j(z_1^+)\, \mq_+\, \oa\big)\,| ({I\setminus \{j\}})\cup\{1\}\rangle\\
  &\quad+\sum_{\substack{j,k\in{I}\\j<k}} \mathcal{C}^{I}_{j,k}(z_1^+)\,\oa\,\mq_+\, \oa\, \vert {({I}\setminus \{j,k\})\cup \{1\}}\rangle\,.
\end{split}  
\end{equation} 
Here the caligraphic coefficients can be found in \eqref{acof}--\eqref{cccof}. As we can see, only the first term can contribute to $\mathcal{F}^{\,\phi}_1$ as all others already contain less than $m-1$ B-operators. When substituting $\mathcal{A}^I$ and $\mathcal{D}^I$ as given in \eqref{acof} and \eqref{dcof} into \eqref{actx} we obtain
\begin{equation}
\begin{split}\label{qev1}
\big(\mathcal{D}^I(z_1^+)\,\oa\, \mq_+(y)&-\mathcal{A}^I(z_1^+)\, \mq_+(y)\, \oa\big)\,\vert {I}\rangle=\prod_{i=2}^{M}(y-z_i^+)\B(z_i^+)\\
&\times\bigg[\delta(z_1^+)\prod_{i=2}^M f(z_1^+,z_i^+)\oa\mqm(y) -\alpha(z_1^+)\prod_{i=2}^M f(z_i^+,z_1^+)\mqm(y) \oa \bigg]\vert\Omega\rangle+\ldots\,.
 \end{split}
\end{equation}
Here the dots denote terms that contribute to $\mathcal{F}_k^{\,\phi}$ with $k>1$ and as such contain less B-operators. 
Finally, we evaluate the trace over the auxiliary oscillator space, cf.~\eqref{qactt} and \eqref{qop}, on the reference state \eqref{traceev} and obtain
\begin{equation}\label{unwantf}
 \mathcal{F}^{\,\phi}_1(y,z_{1}^+,\ldots,z_{m}^+)=e^{+i\phi}\prod_{i=2}^{\m}(y-z_i^+)\big[e^{+i\phi}\delta(z_1^+)\prod_{i=2}^\m f(z_1^+,z_i^+)-e^{-i\phi}\alpha(z_1^+)\prod_{i=2}^\m f(z_i^+,z_1^+)\big]\,.
\end{equation} 
We demand that the coefficients $\mathcal{F}_1^\phi$ vanish for all permutations of the parameters $z_i^+$, cf.~\eqref{cond1}. This yields the Bethe equations
\begin{equation}\label{betheeq}
Q_+(z_k^+-1)\alpha(z_k^+)+Q_+(z_k^++1)\delta(z_k^+)=0\,,\quad \text{for}\quad k=1,\ldots,\m\,,
\end{equation}  
compare \eqref{baein2} and \eqref{ad}. Here we rewrote the product of functions $f$ in \eqref{fg} for $z_i^+\neq z_k^+$ in terms of Baxter's Q-function as given in \eqref{qfunk}. 
\subsection{Ravelled Bethe equations: Recursion}\label{sec:second}
Naturally, we expect that also the coefficients $\mathcal{F}_k^{\,\phi}$ with $k>1$ vanish if the Bethe equations are satisfied. This is indeed the case. However, the coefficients are quite involved and we will not present them here. In order to show that they vanish we again  consider the first term in the sum on the right-hand-side of \eqref{qev1} and isolate the part that does not depend on $\B(z_1^+)$. 
As $\B(z_1^+)$ appears on the right of the monodromy $\mq_+$ in the last two terms of \eqref{actx} we have to keep these terms and evaluate them further using \eqref{actqmon}. We find 
\begin{equation}\label{actxdrop}
\begin{split}
  \X(y,z_1^+)\vert I\rangle&\simeq\big(\mathcal{D}^I(z_1^+)\,\oa\, \mq_+(y)-\mathcal{A}^I(z_1^+)\, \mq_+(y)\, \oa\big)\,\vert {I}\rangle\\
  &\quad+\sum_{j\in {I}}\mathcal{C}^{I}_j(z_1^+)\,\oa\,\mq_+(y)\, \oa\,\vert {{I}\setminus \{j\}}\rangle\\
  &\quad+\sum_{j\in I}\big(\mathcal{D}^I_j(z_1^+)\,\oa\,  \X(y,z_1^+)-\mathcal{A}^I_j(z_1^+)\,  \X(y,z_1^+)\, \oa\big)\,| {I\setminus \{j\}}\rangle\\
  &\quad+\sum_{\substack{j,k\in{I}\\j<k}}\mathcal{C}^{I}_{j,k}(z_1^+)\,\oa\, \X(y,z_1^+)\, \oa\, \vert {{I}\setminus \{j,k\}}\rangle\,,
\end{split}  
\end{equation}  
where we dropped the terms with $\B(z_1^+)$ on the left of the monodromy $\mq_+$.
From \eqref{actxdrop} and \eqref{eqrel} it becomes clear that  the part which does not depend on $\B(z_1^+)$ of $\vert y,\{z^+\}\rangle$ can be written as
\begin{equation}\label{ravel}
\begin{split}
 \vert y,\{z^+\}\rangle\simeq \X(y,z_1^+)\vert I\rangle\simeq \sum_{\sigma\in\mathcal{S}_{m-1}} \sum_{k=1}^{\m}\,\frac{1}{\vert J_k\vert!\,\vert I\setminus J_k\vert!}\,\mathbf{G}^I_{\sigma(J_k)}&(y,z_1^+)\,\vert\sigma(I\setminus J_k)\rangle\,,
 \end{split}
\end{equation} 
where $J_k=\{2,\ldots,k\}$. The coefficients $\mathbf{G}^I_{J_k}$ are symmetric in $z_i^+$ for $i\in I/J_k$ and $z_j^+$ for $j\in J_k$ and depend on the monodromy $\mq_+$ and the oscillators $\oa$ but not on the operators $\A$, $\B$, $\C$ and $\D$. They are determined in the following. Substituting \eqref{ravel} into \eqref{actxdrop} we obtain the recursion relation
\begin{equation}\label{recurs}
\begin{split}
 \mathbf{G}^I_{J_k}(y,z_1^+)&=\sum_{j=2}^k\big(\mathcal{D}^I_j(z_1^+)\,\oa\,  \mathbf{G}^{I\setminus \{j\}}_{J_k\setminus\{j\}}(y,z_1^+)-\mathcal{A}^I_j(z_1^+)\,  \mathbf{G}^{I\setminus \{j\}}_{J_k\setminus\{j\}}(y,z_1^+)\, \oa\big)\\
 &\quad+\sum_{2\leq i<j\leq k}\mathcal{C}^{I}_{i,j}(z_1^+)\,\oa\, \mathbf{G}^{I\setminus \{i,j\}}_{J_k\setminus\{i,j\}}(y,z_1^+)\, \oa\,,
 \end{split}
\end{equation} 
for all permutations of the elements $2,\ldots,\m$ with the initial conditions
\begin{equation}\label{bdkl1}
 \mathbf{G}^I_{\emptyset}(y,z_1^+)=\mathcal{D}^I(z_1^+)\,\oa\, \mq_+(y)-\mathcal{A}^I(z_1^+)\, \mq_+(y)\, \oa\,,
\end{equation} 
\begin{equation}\label{bdkl}
\begin{split}
 \mathbf{G}^I_{\{i\}}(y,z_1^+)=\mathcal{D}^{I}_i(z_1^+)\mathcal{D}^{I\setminus \{i\}}(z_1^+)\,\oa^2\, \mq_+(y)+\mathcal{A}^{I}_i(z_1^+)\mathcal{A}^{I\setminus \{i\}}(z_1^+)\, \mq_+(y)\, \oa^2\,.
\end{split}
\end{equation} 
Note that in \eqref{bdkl} we expressed $\mathcal{C}_j$ in terms of $\mathcal{A}$ and $\mathcal{D}$ using \eqref{ccof}. 
The operators $\mathbf{G}^I_{J_k}$ are obtained recursively in Appendix~\ref{sec:proofrec}. If we substitute $\mathcal{A}$ and $\mathcal{D}$ as given in \eqref{acof} and \eqref{dcof} they read
\begin{equation}\label{gkk}
\begin{split}
\mathbf{G}_{J}^I(y,z_1^+,z_I^+)=\vert J\vert!\prod_{i\in J}g(z_i^+,z_1^+)\bigg[&\,\oa^{\vert J\vert+1}\, \mq_+(y)\prod_{j\in J\cup\{1\}}\delta(z_j^+)\prod_{l\in I\setminus J}  f(z_j^+,z_l^+)\\&
-\mq_+(y)\,\oa^{\vert J\vert+1}\prod_{j\in J\cup\{1\}}\alpha(z_j^+)\prod_{l\in I\setminus J}f(z_l^+,z_j^+)\bigg]\,.
\end{split}
\end{equation} 
In principle, one can evaluate the coefficients $\mathcal{F}_k^{\,\phi}$ by subsequently applying \eqref{ravel} and taking the trace as in \eqref{unwanq}. However, this is not necessary! Taking the trace of \eqref{ravel} and moving all oscillators to the right of the monodromy $\mq_+$ yields
\begin{equation}\label{wurst}
\begin{split}
 \vert y,\{z^+\}\rangle_\phi\simeq\sum_{\sigma\in\mathcal{S}_{m-1}} \sum_{k=1}^{\m}&\frac{e^{i(y+k)\phi}\prod_{i=2}^{k}g(z_{\sigma(i)}^+,z_1^+)}{(\m-k)!Z_+}
 \mathcal{G}_k^\phi(y,z_1^+,z_{\sigma(2)}^+,\ldots,z_{\sigma(\m)}^+)\\
 &\times\tr\big[ e^{-2i\phi\oad\oa}\mq_+(y)\oa^{k}\big]
\,\B(z_{\sigma(k+1)}^+)\cdots \B(z_{\sigma(\m)}^+)\vert\Omega^+\rangle\,,
\end{split}
\end{equation} 
with the coefficients
\begin{equation}\label{gcof}
 \mathcal{G}_k^\phi(y,z_1^+,z_2^+,\ldots,z_\m^+)=e^{+ik\phi}\prod_{j=1}^{k}\delta(z_j^+)\prod_{l=k+1}^\m  f(z_j^+,z_l^+)-e^{-ik\phi}\prod_{j=1}^{k}\alpha(z_j^+)\prod_{l=k+1}^\m f(z_l^+,z_j^+)\,.
\end{equation} 
We note that up to a factor  $\mathcal{G}_1^\phi$ coincides with $\mathcal{F}_1^{\,\phi}$ and thus vanishes.
To show that all  $\mathcal{G}_k^{\,\phi}$ with $k>1$ vanish  we note that the Bethe equations \eqref{betheeq} imply
\begin{equation}\label{baerave}
e^{+i\vert I\vert\phi}\prod_{j\in I}\delta(z_j^+)\prod_{l\in\bar I}  (z_j^+-z_l^++1)-e^{-i\vert I\vert\phi}\prod_{j\in I}\alpha(z_j^+)\prod_{l\in \bar I} (z_j^+-z_l^+-1)=0\,,
\end{equation} 
where we introduced the set $I\subseteq\{1,\ldots,\m\}$ of cardinality $\vert I\vert$ and its complement $\bar I$. 
From \eqref{baerave} it follows that
\begin{equation}
 \mathcal{G}_k^{\,\phi}(y,z_1,z_{\sigma(2)},\ldots,z_{\sigma(\m)})=0\,.
\end{equation} 
As a consequence all unwanted terms that are independent of $\B(z_1^+)$ vanish, cf.~\eqref{wurst}. 

We discussed earlier that the unwanted terms and in particular \eqref{unwanmon} are invariant under permutations of the Bethe roots. Following the same strategy presented here for the case not containing $\B(z_1^+)$ we can equally well evaluate the terms that do not contain $\B(z_2^+)$, $\B(z_3^+)$, etc. The corresponding coefficients $\mathcal{G}_k^\phi$ are permutations of \eqref{gcof} involving $z_1^+$ and thus vanish due to \eqref{baerave}. Finally, we conclude that in general the unwanted terms \eqref{coeffs} vanish if the Bethe equations \eqref{betheeq} are satisfied.

\section{The other Q}
To diagonalize the Q-operator $\qop_-$ we can in principle repeat the procedure explained in Section~\ref{fq2q} using the commutation relations in Appendix~\ref{fcrsq2}. The natural choice of the reference state is such that it is annihilated by the B-operator and excitations are created by acting with C-operators, see below. However, for finite-dimensional representations we can make use of the \textit{spin-flip symmetry} of the Q-operators. First, we note that Lax operators $\mathcal{R}_{s,\pm}$ can be related to each other by the transformation
\begin{equation}\label{roprel}
 K\mathcal{R}_{s,\pm}(y)K^{-1}=\mathcal{R}_{s,\mp}(y)\,,
\end{equation} 
defined as
\begin{equation}
 K\, S_\pm\, K^{-1}=S_\mp\,,\quad\quad  K\,S_3\,K^{-1}=-S_3\,.
\end{equation} 
As a consequence of \eqref{roprel} and the definition of the Q-operators in \eqref{qop} we see that $\qop_\pm$ are related through the transformation $K$ acting on each site of the spin chain and a sign change of the twist field $\phi$
\begin{equation}\label{qsymm}
\mathbf{K}\qop_\pm(y)\mathbf{K}^{-1}=\qop_\mp(y)\vert_{\phi\,\rightarrow\, -\phi}\quad\quad\text{with}\quad\quad  \mathbf{K}=\underbrace{K\otimes K \otimes\ldots \otimes K}_N.
\end{equation}  
As shown in Section~\ref{fq2q} the Q-operator $\qop_+$ acts diagonally on the Bethe states $\vert\psi_\m^+ \rangle$ if the Bethe equations \eqref{betheeq} are satisfied. 
As a consequence of \eqref{qsymm} we obtain the action of the Q-operator $\qop_-$ on the Bethe states
\begin{equation}\label{bstat}
 \vert\psi_\m ^-\rangle=\C(z^-_1)\ldots \C(z^-_\m )\vert\Omega^-\rangle \,,
\end{equation} 
with $\vert\Omega^-\rangle =\vert\omega^-\otimes\ldots\otimes\vert\omega^-\rangle=K\vert\Omega^+\rangle$.
The Q-operator $\qop_-$ acts diagonally on the Bethe states \eqref{bstat} as
\begin{equation}
 \qop_-(y)\vert \psi^-_\m \rangle =e^{- i y\phi}\prod_{k=1}^{m}(y-z^-_k)\vert \psi^-_\m \rangle\,,
\end{equation} 
if the Bethe roots $z_i^-$ satisfy the Bethe equations 
\begin{equation}
Q_-(z^-_k-1)\alpha(z^-_k)+Q_-(z^-_k+1)\delta(z^-_k)=0\,,
\end{equation} 
with $\alpha$ and $\delta$ given in \eqref{ad}, cf.~\eqref{qpsi}--\eqref{baein2}. Here we used that the transformation $\mathbf{K}$ acts on the quantum space of the monodromy $\mathcal{M}$ in \eqref{transmon} as
\begin{equation}
 \mathbf{K} \left(\begin{array}{cc}
              \A(y)&\B(y)\\
	      \C(y)&\D(y)
             \end{array}\right) \mathbf{K}^{-1}=\left(\begin{array}{cc}
              \D(y)&\C(y)\\
	      \B(y)&\A(y)
             \end{array}\right)\,.
\end{equation} 

\section{Conclusion}
We diagonalized the Q-operators for the $\text{XXX}_s$ Heisenberg spin chain using the algebraic Bethe ansatz. 
In doing so we demonstrate that the eigenvalues of the Q-operators are given by Baxter's Q-functions.
In analogy to the algebraic Bethe ansatz for the corresponding transfer matrix, the results obtained rely on certain fundamental commutation relations that were derived from the Yang-Baxter equation and the cancellation of the so-called unwanted terms.
 Our findings consolidate the construction of the Q-operators as carried out in~\cite{Bazhanov2010a,Frassek2011}. Here the functional form of the Q-operators has been deduced rather indirectly from their functional relations.

 We have seen that the commutation relations derived here are convenient to diagonalize the Q-operators $\qop_\pm$ on the states $\vert\psi^\pm\rangle$ respectively as the eigenvalues are manifestly polynomials. However, it is not apparent that the eigenvalue will be a polynomials when acting with  $\qop_\pm$ on the states $\vert\psi^\mp\rangle$ respectively. This yields a curious relation among the Bethe roots as presented in Appendix~\ref{dualroot}, see also~\cite{PronkoJ.Phys.A32:2333-23401999,Korff2006}. It may be worth to study these relations in the zero twist limit and in particular at half-filling.

Furthermore, it would be interesting to generalize our procedure to the case of $\mathfrak{sl}(n)$-invariant spin chains and their supersymmetric relatives~\cite{Bazhanov2010,Frassek2011,Frassek2010} where the nested Bethe ansatz has to be applied. 

Finally, we hope that our proof carries over to the $q$-deformed case where a similar procedure has been studied in~\cite{Korff2004}.

 \section*{Acknowledgments}
I like to that Patrick Dorey, Frank G\"ohmann and Matthias Staudacher for inspiring discussions and comments on the manuscript. Istv\'{a}n Sz\'{e}cs\'{e}nyi for his interest in the problem and various fruitful discussions.  Furthermore, I thank Zolt\'{a}n Bajnok and Rafael Nepomechie for interesting comments on my talk during the workshop on finite-size technology in low dimensional quantum systems in Budapest, June 2014. 

The research leading to these results has received funding from the People Programme
(Marie Curie Actions) of the European Union’s Seventh Framework Programme FP7/2007-
2013/ under REA Grant Agreement No 317089 (GATIS).

\section*{Appendices}
\appendix 
\section{Fundamental commutation relations for \texorpdfstring{$\mathbf{Q}_-$}{}}\label{fcrsq2}
The fundamental commutation relations arising from the Yang-Baxter equation \eqref{fcrrtt} for the Q-operator $\qop_-$ read
\begin{gather}
 \mq_-(y)[(x-y+1+\oad\oa)\D(x)+\oad\,\C(x)]=[\D(x)(x-y+1+\oad\oa)-\B(x)\,\oa ]\mq_-(y)\,,\\
 \nonumber\\
  \mq_-(y)\C(x)=\C(x)(y-x-1-\oad\oa)\mq_-(y)+\A(x)\,\oa\, \mq_-(y)-\mq_-(y)\,\oa\,\D(x)\,,\label{mbaa}\\
  \nonumber\\
   \B(x)\mq_-(y)=\mq_-(y)(y-x-1-\oad\oa)\B(x)+\D(x)\,\oad\, \mq_-(y)-\mq_-(y)\,\oad\,\A(x)\,,\\
   \nonumber\\
    [\A(x),\mq_-(y)]=\C(x)\,\oad\, \mq_-(y)+\mq_-(y)\,\oa\, \B(x)\,.\label{bdaa}
\end{gather} 
Combining \eqref{mbaa} and \eqref{bdaa} yields
\begin{equation}\label{combinedaa}
 \mq_-(y)\C(x)=(y-x)\C(x)\mq_-(y)+\oa \mq_-(y) \A(x)+\oa \mq_-(y)\oa \B(x)-\mq_-(y)\oa\D(x)\,.
\end{equation} 
\section{Commutation relations}\label{comrel}
In this appendix we calculate the action of the operators $\A$, $\D$ and $\C$ on a given off-shell state
\begin{equation}
 \vert {I}\rangle=\prod_{k\in I}\B(z_{k})\vert \Omega \rangle\,.
\end{equation} 
We are mostly interested in the case where the set $I$ is given by $I=\{2,3,\ldots,\m\}$. One finds
\begin{gather}
 \A(z_1)\vert {I}\rangle=\mathcal{A}^I(z_1)\vert {I}\rangle+\sum_{j\in I}\mathcal{A}^I_j(z_1)  \big| ({I\setminus \{j\}})\cup\{1\}\big\rangle\,,\label{actaa}\\
 \nonumber\\
 \D(z_1)\vert {I}\rangle=\mathcal{D}^I(z_1)\vert {I}\rangle+\sum_{j\in I}\mathcal{D}^I_j(z_1)  \big| ({I\setminus \{j\}})\cup\{1\}\big\rangle\,,\label{actbb}\\
  \nonumber\\
    \C(z_1)\vert {I} \rangle=\sum_{j\in {I}}\mathcal{C}^{I}_j(z_1)\vert {{I}\setminus \{j\}}\rangle+\sum_{\substack{j,k\in{I}\\j<k}} \mathcal{C}^{I}_{j,k}(z_1) \vert {({I}\setminus \{j,k\})\cup \{1\}}\rangle\,.\label{actcc}
\end{gather} 
The coefficients $\mathcal{A}$, $\mathcal{D}$ and $\mathcal{C}$ appearing above read
\begin{equation}\label{acof}
 \mathcal{A}^I(z_1)=\alpha(z_1)\prod_{k\in I}f(z_k,z_1)\,,\quad  \mathcal{A}^I_j(z_1)=\alpha(z_j)\,g(z_1,z_j)\prod_{\substack{k\in{I}\setminus \{j\}}}f(z_k,z_j)\,,
\end{equation} 
\begin{equation}\label{dcof}
 \mathcal{D}^I(z_1)=\delta(z_1)\prod_{k\in I}f(z_1,z_k)\,,\quad  \mathcal{D}^I_j(z_1)=\delta(z_j)\,g(z_j,z_1)\prod_{\substack{k\in{I}\setminus \{j\}}}f(z_j,z_k)\,,
\end{equation} 
and
 \begin{equation}\label{ccof}
 \begin{split}
   \mathcal{C}^{I}_j(z_1)=\mathcal{A}_j^I(z_1)\,\mathcal{D}^{I\setminus j}(z_1)+\mathcal{D}_j^I(z_1)\,\mathcal{A}^{I\setminus j}(z_1)\,,
 \end{split}
\end{equation} 
\begin{equation}\label{cccof}
 \begin{split}
\mathcal{C}^{I}_{j,k}(z_1)&=\mathcal{A}^{I}_j(z_1)\,\mathcal{D}_k^{I\setminus j}(z_1)+\mathcal{D}_j^I(z_1)\,\mathcal{A}^{I\setminus j}_k(z_1)\,.
 \end{split}
\end{equation} 
The formulas above can be deduced from the commutation relations in \eqref{abcom} and \eqref{dbcom}, see also~\cite{Slavnov2007} for a more detailed explanation. 
\section{Proof of the recursion relation}
\label{sec:proofrec}
Let us first note that \eqref{gkk} is compatible with the initial conditions \eqref{bdkl1} and \eqref{bdkl}. In order to show that $\mathbf{G}_{J}^I$ \eqref{gkk} solves the recursion relation \eqref{recurs} we write \eqref{gkk} as
\begin{equation}
\begin{split}\label{defg}
\mathbf{G}_{J}^I(y,z_1^+)=\mathbf{D}_{J}^I(y,z_1^+)-\mathbf{A}_{J}^I(y,z_1^+)\,.
\end{split}
\end{equation} 
Here we defined
\begin{equation}
\begin{split}
\mathbf{D}_{J}^I(y,z_1^+)=\vert J\vert!\prod_{i\in J}g(z_i^+,z_1^+)\prod_{j\in J\cup\{1\}}\delta(z_j^+)\prod_{l\in I\setminus J}  f(z_j^+,z_l^+)\,\oa^{\vert J\vert+1}\, \mq_+(y)\,,
\end{split}
\end{equation} 
and
\begin{equation}
\begin{split}
\mathbf{A}_{J}^I(y,z_1^+)=\vert J\vert!\prod_{i\in J}g(z_i^+,z_1^+)\prod_{j\in J\cup\{1\}}\alpha(z_j^+)\prod_{l\in I\setminus J}  f(z_l^+,z_j^+)\, \mq_+(y)\,\oa^{\vert J\vert+1}\,.
\end{split}
\end{equation}
Our proof is based on two identities
\begin{equation}\label{recursii}
\begin{split}
 \mathbf{G}^I_{J_k}(y,z_1^+)&=\mathbf{D}_{J_k}^I(y,z_1^+)-\mathbf{A}_{J_k}^I(y,z_1^+)\\
 &=\sum_{j=2}^k\big(\mathcal{D}^I_j(z_1^+)\,\oa\,  \mathbf{D}^{I\setminus \{j\}}_{J_k\setminus\{j\}}(y,z_1^+)+\mathcal{A}^I_j(z_1^+)\,  \mathbf{A}^{I\setminus \{j\}}_{J_k\setminus\{j\}}(y,z_1^+)\, \oa\big)\,,
\end{split}
\end{equation} 
and
\begin{equation}\label{recursi}
\begin{split}
\sum_{2\leq i<j\leq k}\mathcal{C}^{I}_{i,j}(z_1^+)\,\oa\, \mathbf{G}^{I\setminus \{i,j\}}_{J_k\setminus\{i,j\}}(y,z_1^+)\, \oa= \sum_{j=2}^k\big(&\mathcal{D}^I_j(z_1^+)\,\oa\,  \mathbf{A}^{I\setminus \{j\}}_{J_k\setminus\{j\}}(y,z_1^+)\\
&+\mathcal{A}^I_j(z_1^+)\,  \mathbf{D}^{I\setminus \{j\}}_{J_k\setminus\{j\}}(y,z_1^+)\, \oa\big)\,,
 \end{split}
\end{equation}
Substituting \eqref{defg} for $\mathbf{G}_{J_k\setminus\{j\}}^{I\setminus\{j\}}$ in the recursion relation \eqref{recurs} we see that if \eqref{recursii} and \eqref{recursi} hold also \eqref{recurs} is satisfied.

To show \eqref{recursii} and \eqref{recursi} we note that
\begin{equation}\label{dfun}
 \mathbf{D}_{J}^I(y,z_1^+)=+\sum_{j\in J}\mathcal{D}^I_j(z_1^+)\,\oa\,  \mathbf{D}^{I\setminus \{j\}}_{J\setminus\{j\}}(y,z_1^+)\,,
\end{equation} 
\begin{equation}\label{afun}
\mathbf{A}_{J}^I(y,z_1^+)=-\sum_{j\in J}\mathcal{A}^I_j(z_1^+)\,  \mathbf{A}^{I\setminus \{j\}}_{J\setminus\{j\}}(y,z_1^+)\,\oa\,,
\end{equation} 
with the coefficients $\mathcal{A}$ and $\mathcal{D}$ given in \eqref{acof} and \eqref{dcof}. These relations rely on the curious identity
\begin{equation}
 \sum_{k=1}^p\prod_{i\neq k}^p f(x_i,x_k)=p\,,
\end{equation} 
which generalizes to the trigonometric case and can be shown by induction in $p\in\mathbb{N}$ and using the partial fraction decomposition 
\begin{equation}
 \prod_{i=1}^p\frac{x-x_i-1}{x-x_i}=1-\sum_{k=1}^p\frac{1}{x-x_k} \prod_{i\neq k}\frac{x_k-x_i-1}{x_k-x_i}\,,
\end{equation} 
It immediately follows that \eqref{recursii} holds.
Furthermore, we note that the term on the left hand side of \eqref{recursi} can be written as
\begin{equation}\label{recurswi}
\begin{split}
\sum_{2\leq i<j\leq k}\mathcal{C}^{I}_{i,j}(z_1^+)\,\oa\, \mathbf{G}^{I\setminus \{i,j\}}_{J_k\setminus\{i,j\}}(y,z_1^+)\, \oa&
=\sum_{i=2}^k\mathcal{A}^{I}_i(z_1)\,\sum_{\substack{j=2\\j\neq i}}^k\mathcal{D}_j^{I\setminus i}(z_1)\,\oa\, \mathbf{D}^{I\setminus \{i,j\}}_{J_k\setminus\{i,j\}}(y,z_1^+)\\&
\quad-\sum_{i=2}^k\mathcal{D}^{I}_i(z_1)\,\sum_{\substack{j=2\\j\neq i}}^k\mathcal{A}_j^{I\setminus i}(z_1)\,\mathbf{A}^{I\setminus \{i,j\}}_{J_k\setminus\{i,j\}}(y,z_1^+)\, \oa\,,
 \end{split}
\end{equation}
see Appendix~\ref{comrel}. Again using \eqref{dfun} and \eqref{afun} we arrive at \eqref{recursi} which concludes the proof.

\section{Hamiltonian}
For completeness we give the Hamiltonian for the corresponding $\text{XXX}_\frac{1}{2}$ Heisenberg spin chain. It can be obtained from the transfer matrix in \eqref{transi} by taking the logarithmic derivative at the shift-point, see e.g.~\cite{Faddeev2007}. It reads
\begin{equation}
 \mathbf{H}=2\sum_{i=1}^N\left(\mathbb{I}-\mathbf{P}_{i,i+1}\right)\,,
\end{equation} 
with the permutation operator $\mathbf{P}_{i,i+1}=(\mathbb{I}+4\vec{S}^{(i)}\vec{S}^{(i+1)})/2$ and  the twisted boundary conditions
\begin{equation}
 \mathbf{P}_{N,N+1}=e^{-2i\phi S_3^{(N)}}\,\mathbf{P}_{N,1}\,e^{+2i\phi S_3^{(N)}}\,.
\end{equation} 
Here the generators $\vec{S}^{(i)}=\frac{1}{2}\vec{\sigma}^{(i)}$ at site $i$ are given in terms of the Pauli matrices $\vec{\sigma}$. 
\section{Relations between both sides of the equator}\label{dualroot}
Solutions beyond the equator were studied in~\cite{PronkoJ.Phys.A32:2333-23401999}. Here we present a relation between the roots $z_i^+$ and $z_j^-$ as also obtained in~\cite{Korff2006} for $s=1/2$. The derivation relies on the property that $\qop_\pm$ have common eigenvectors and the fundamental commutation relations \eqref{amm} and \eqref{cmm}. For $s=1/2$ we find
\begin{equation}
 \sum_{q=0}^\infty e^{-2i\phi q}(y-q-\frac{1}{2})^N\prod_{i=1}^\m\frac{(y-z_i^-)}{(y-z_i^--q)(y-z_i^--1-q)}=\prod_{i=1}^{N-\m}(y-z_i^+)\,.
\end{equation} 
We can set $y=z_k^+$ and obtain the relations
\begin{equation}
 \sum_{q=0}^\infty e^{-2i\phi q}\big(z_k^+-q-\frac{1}{2}\big)^N\prod_{i=1}^\m\frac{(z_k^+-z_i^-)}{(z_k^+-z_i^--q)(z_k^+-z_i^--1-q)}=0\quad\text{for}\quad k=1,\ldots,N-\m\,.
\end{equation} 
This formula has been checked for a few examples. Similar relations hold for finite-dimensional representations of higher spin. One finds
\begin{equation}
\sum_{q=0}^\infty e^{-2i\phi q}\,\mq_+^{\,\Omega^-}(y,s,q)\prod_{i=1}^\m\frac{(y-z_i^-)}{(y-z_i^--q)(y-z_i^--1-q)}=\prod_{i=1}^{2sN-\m}(y-z_i^+)\,,
\end{equation} 
with the action of the diagonal elements of the monodromy on the reference state $\vert\Omega^-\rangle$ given by
\begin{equation}
 \mq_+^{\,\Omega^-}(y,s,q)=\left(\frac{\Gamma(y+s)}{\Gamma(y-s)}\,_2F_1(-2s,-q,1-s-y,1)\right)^N\,,
\end{equation} 
which can be expressed in terms of $\Gamma$-functions.
\bibliographystyle{hieeetr}
\bibliography{qref}{}

\end{document}